\documentclass[twocolumn,prl]{revtex4}
\usepackage{epsfig,amssymb,amsmath}
\begin{document}
\title{Charging of Superconducting Layers and Novel Type of Hysteresis in Coupled Josephson Junctions}

\author{ Yu. M. Shukrinov~$^{1}$}
\author{ M. A. Gaafar~$^{1,2}$}
\address{$^{1}$ BLTP, JINR, Dubna, Moscow Region, 141980, Russia \\
$^{2}$Department of Physics, Faculty of Science, Menoufiya University, Egypt.}

\begin{abstract}

A manifestation of a novel type of hysteresis related to the parametric resonance in the system of coupled
Josephson junctions is demonstrated. Opposite to McCumber and Steward hysteresis, we find that the width of this
hysteresis is inversely proportional to the McCumber parameter and depends also on coupling between junctions
and the boundary conditions. An investigation of time dependence of the electric charge in superconducting
layers allow us to explain the origin of this hysteresis by different charge dynamics for increasing and
decreasing bias current processes. The effect of wavelength of the longitudinal plasma waves created at the
resonance  on the charging of superconducting layers  is demonstrated.  We found a strong effect of the
dissipation in the system on the amplitude of the charge oscillations at the resonance.
\end{abstract}

\maketitle The hysteresis features of single Josephson junction (JJ) were studied by McCumber and Steward long
time ago.~\cite{mccumber68,steward68} Particularly, it was shown that the width of the hysteresis is directly
proportional to the McCumber parameter $\beta_c$ related to the dissipation parameter $\beta$ by $\beta_c=
1/\beta^2$. In compare with the single Josephson junction, the system of the coupled Josephson junctions has a
multiple branch structure and the definition of the return current is more general now: the system can return to
the zeroth voltage state from any branch. The branches have a breakpoint (BP) and a breakpoint region (BPR)
before transition to another branch.~\cite{sm-prl07} The BP current characterizes the resonance point at which
the longitudinal plasma wave (LPW) with a definite wave number k is created by Josephson oscillations in stacks.
Hysteresis features of intrinsic JJ in high temperature superconductors have a wide interest also due to the
observed powerful coherent radiation from such system.~\cite{Ozyuzer} In Ref.~\cite{koyama09} the authors
summarized the experimental results and stressed that the strong emission was observed near the unstable point
of the return current in the uniform voltage branch.  The radiation is related to the same region in the current
voltage characteristics (CVC) where the BP and the BPR were observed and it made the phase dynamics
investigation of intrinsic JJ corresponding to these parts of CVC an actual problem today.

Since thickness of superconducting layer (S-layer) in the intrinsic JJ is comparable to the Debye screening
length, the S-layers are in the nonstationary nonequilibrium state due to the injection of quasiparticle and
Cooper pairs.~\cite{koyama96,ryndyk98}. The charge neutrality in S-layers is locally broken and this charging
effect modifies the Josephson relation between voltage and the phase difference. The question concerning the
value of the electric charge in S-layer and its maximum realized at the parametric resonance is not investigated
yet.

In this paper we study the phase dynamics in coupled Josephson junctions. A novel type of hysteresis related to
the parametric resonance in this system is demonstrated. We show that the width of this hysteresis is inversely
proportional to the McCumber parameter and depends on coupling parameter of the system and the boundary
conditions. The origin of this hysteresis is related to the different charge dynamics for increasing and
decreasing bias current processes. We discuss the question concerning the maximal electric charge  in S-layers
realized at the resonance and show that it depends on the relation between the wavelength of LPW and the period
of lattice. We demonstrate a strong effect of the dissipation in the system on the coefficient of the
exponential growth of the maximal electric charge in S-layers.

First we discuss the phase dynamics in the system of coupled JJ. The CVC of JJs are numerically calculated in
the framework of capacitively coupled Josephson junction model with diffusion current (CCJJ+DC)
~\cite{sms-physC06}. The system of equations for the gauge-invariant phase differences $\varphi_l(\tau)=
\theta_{l+1}(\tau)-\theta_{l}(\tau)-\frac{2e}{\hbar}\int^{l+1}_{l}dz A_{z}(z,\tau)$ between $S$-layers in this
model has the form $\frac{d^2}{dt^2}\varphi_{l}=(1-\alpha\nabla^{(2)})(I-\sin(\varphi_{l})
-\beta\frac{d\varphi_{l}}{dt})$ where $\theta_{l}$ is the phase of the order parameter in the S-layer $l$, $A_z$
is the vector potential in the barrier, $\alpha$ and $\beta$ are coupling and dissipation parameters,
respectively, and $\nabla^{(2)}f_l=f_{l+1}+f_{l-1}-2f_{l}$. In our simulations we use both periodic and
nonperiodic boundary conditions (BC). At nonperiodic BC it is suggested that the first and the last S-layers are
in contact with normal metals and their effective width $s_0$ and $s_N$ may be extended due to the proximity
effect into attached metals. Nonperiodic BC are characterized by parameter $\gamma=s/s_0=s/s_N$ and the
equations for the first and last layers in the system for  phase differences are different from the equations
for the middle S-layers\cite{koyama96,matsumoto99}. We solve the system of dynamical equations for  phase
differences using the fourth order Rungge-Kutta method. We use a dimensionless time $\tau = t\omega_p$, where
$\omega_{p}$ is the plasma frequency $\omega_{p}=\sqrt{2eI_c/\hbar C}$, ${I_c}$ is the critical current and $C$
is the capacitance. In our simulations we measure the voltage in units of $V_0=\hbar\omega_p/(2e)$ and the
current in units of the $I_c$. The details concerning numerical procedure are given in Ref. 9.
\begin{figure}[!ht]
 \centering
\includegraphics[height=60mm]{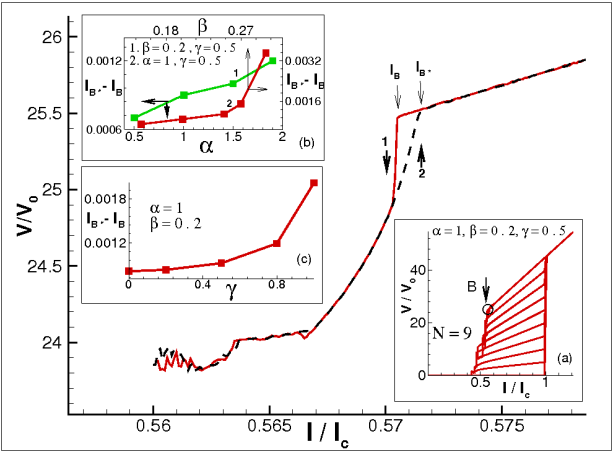}
\caption{Demonstration of the hysteresis behavior in the parametric resonance region in the outermost branch of
the CVC for the stack with 9 JJ at $\alpha= 1$, $\beta= 0.2 $ and $\gamma= 0.5$ . The arrows show the direction
of the bias current sweeping.  The inset (a) shows the total branch structure in the CVC of this stack and BP
location. The inset (b) shows a change of the hysteresis width  $I_{B*}-I_{B}$ with dissipation parameter
$\beta$ and coupling parameter $\alpha$. The inset (c) shows the same with variation of the nonperiodic
parameter $\gamma$.}
 \label{1}
\end{figure}

The results of simulation of CVC and its features near parametric resonance region are presented in
Fig.~\ref{1}. The inset (a) shows the total branch structure of the CVC for the stack with 9 JJ at $\alpha= 1$,
$\beta= 0.2 $ and $\gamma= 0.5$.  The CVC shows the following features: (i) a jump at $I/I_c=1.0$  from the zero
voltage branch to the outermost branch with all junctions in the rotating state; (ii) practically linear
dependence of the voltage on bias current at $I>I_c$; (iii) multiple branching in the hysteresis region. The
circle and arrow with letter $B$ show the BP location on the outermost branch.

Fig.~\ref{1} demonstrates a hysteresis  in the outermost branch of the CVC. It is obtained by decreasing the
bias current till some point in the BPR (curve 1), then we increase the current to pass the resonance region
(curve 2). The arrows show direction of the bias current changing. The hysteresis is characterized by its width
$I_{B*}-I_{B}$, where $I_{B}$ is the value of breakpoint current in decreasing current process and $I_{B*}$ is a
characteristic current value in the increasing current process.

The dependence of the hysteresis width on the dissipation parameter $\beta$ is shown in the inset (b). The width
is increasing with parameter $\beta$ (i.e., it is decreasing with the McCumber parameter). As we mentioned
above, the McCumber and Steward hysteresis  demonstrates an opposite behavior for single JJ. They obtained that
the return current $I_r$ (which characterizes the value of hysteresis) decreases with increasing of the McCumber
parameter.\cite{sm-physC06} So, we have observed  \emph{a novel type of hysteresis related to the parametric
resonance in coupled JJ}. The insets (b) and (s) show as well an increase of the hysteresis width $I_{B*}-I_{B}$
with the coupling parameter $\alpha$ and parameter of nonperiodicity $\gamma$, respectively.

To make clear the origin of this hysteresis, we study time dependence of the charge oscillations in the
S-layers. Using Maxwell equation $\emph{div} (\varepsilon\varepsilon_0 E) = $Q$ $, where $\varepsilon$ and
$\varepsilon_0$ are relative dielectric and electric constants, we express the charge density $Q_l=Q_0 \alpha
(V_{l+1}-V_{l})$ in the S-layer $l$ by the voltages $V_{l}$ and $V_{l+1}$ in the neighbor insulating layers,
where $Q_0 = \varepsilon \varepsilon _0 V_0/r_D^2$,  and $r_D$ is Debay screening length. The charge dynamics in
the S-layers determines the features of current voltage characteristics of the coupled Josephson junctions.
\begin{figure}[ht]
 \centering
\includegraphics[height=60mm]{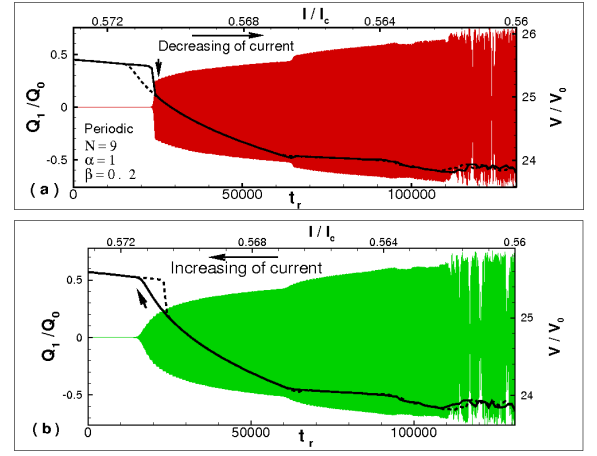}
\caption{Difference in  the charge-time dependence and CVC in (a) decreasing current process and (b) increasing
current process. The thick curves show the CVC.}
 \label{2}
\end{figure}

Solution of the system of dynamical equations for  phase differences gives us the voltages as a functions of
time $V_{l}$(t) in all junctions in the stack, and it allows us to investigate the time dependence of the charge
in each S-layer. Here we investigate the charge-time dependence for two processes: decreasing (Figs. 2(a), 3, 4,
5) and increasing (Fig.~\ref{2}(b)) the bias current through the stack. The recorded time is calculated as $t_r=
t\omega_p +T_m(I_0 -I)/\delta I$ for decreasing bias current process. For increasing current process
(Fig.~\ref{2}(b)) we record the time dependence at bias current value $I$ during the time interval ($t_r,
t_r-T_m$). We put  mostly $T_m=1000$, $\delta \tau=0.05$ and $\delta I=0.0001$ in our simulations.

In Fig.~\ref{2} the time dependence of the charge in  S-layer of the stack with 9 coupled JJ at $\alpha=1$,
$\beta= 0.2 $ and periodic BC is combined with the CVC of the outermost branch. Here and further we present the
charge oscillations in the first S-layer of the stack: the features of the charge oscillations we are interested
in this papers are the same in the all other layers.  Fig.~\ref{2}(a) shows the charge-time dependence when the
current approaches the resonance point in decreasing current process, while Fig.~\ref{2}(b) presents it when the
bias current is increased. We can see that the charge on the S-layer in Fig.~\ref{2}(b) disappears at different
value of current in compare with the current value in decreasing process (Fig.~\ref{2}(a)). We consider that it
is an origin of the observed hysteresis, which is related to the parametric resonance in this system.
\begin{figure}[ht]
 \centering
\includegraphics[height=55mm]{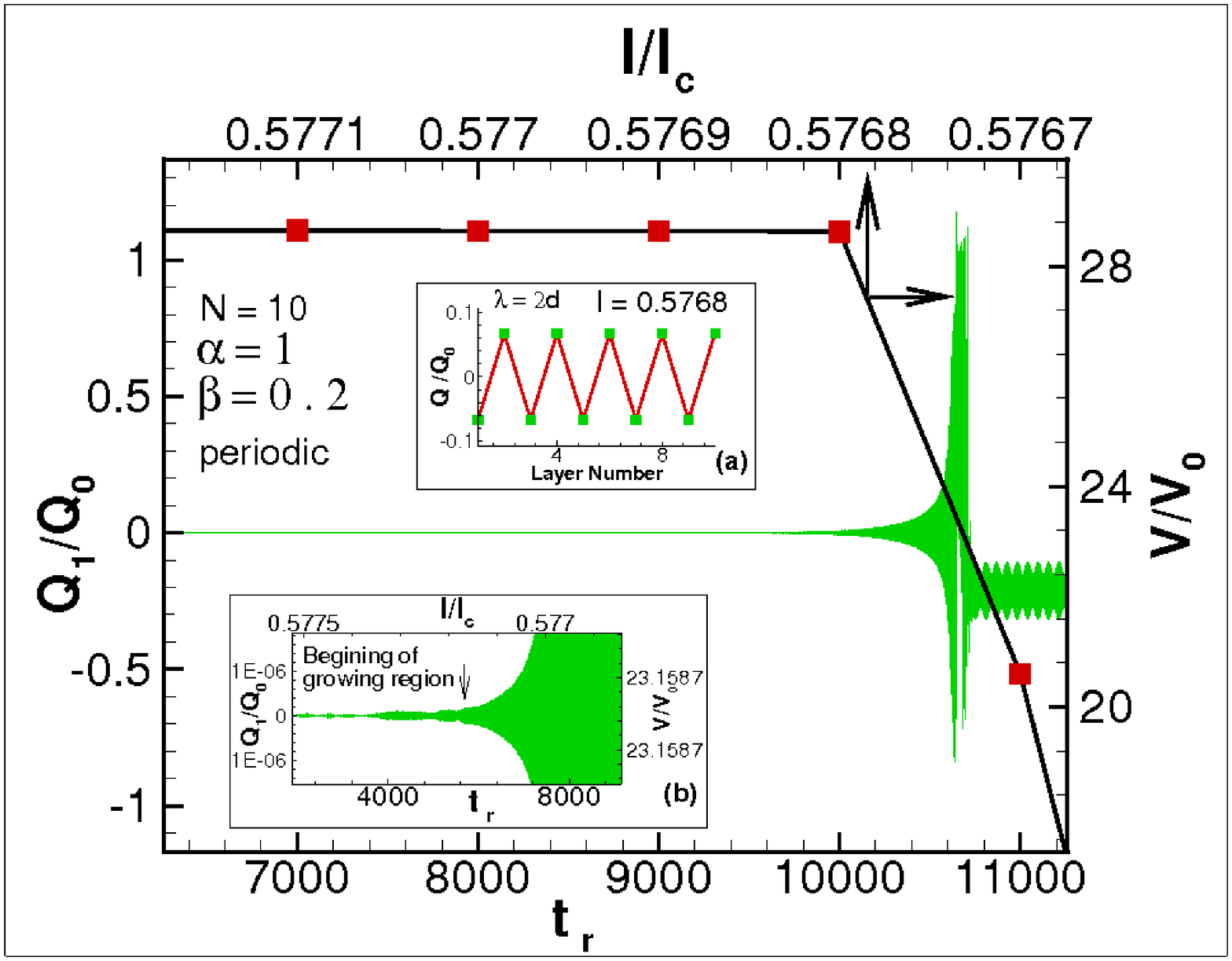}
\caption{Demonstration of the absence of fine structure in charge-time dependence and CVC for the stack with 10
junctions at $\alpha= 1$, $\beta= 0.2$ and periodic BC. The filled squares mark the bias current's steps in CVC.
Inset (a) shows the charge distribution among the layers. Inset (b) illustrates the beginning of the growing
region of the charge in the S-layer.}
 \label{3}
\end{figure}

Let us now discuss the question concerning the amplitude of the electric charge  oscillations in S-layer at
parametric resonance (the parametric resonance corresponds to the BP on the outermost branch of CVC). \emph{Does
its maximal value depend on the wavelength of created LPW?} The Josephson oscillations excite the LPW with
$k=\pi/d$ ($\pi$-mode, wavelength $\lambda=2d$) at the parametric resonance in the stack with even number of JJ
at $\alpha= 1$, $\beta= 0.2$ and periodic BC.\cite{sm-sust07} But in the stacks with odd number $N$ of JJ the
wave number depends on $N$ and it is equal to $k=\pi (N-1)/d N$. In the stacks with even number of JJ the
resonance is "pure", i.e., no additional fine structure appears in CVC.\cite{sms-prb08} So, it is interesting to
compare the maximal value of the electric charge realized in S-layers in this case with the case $\lambda\neq
nd$, where n is integer number.

In Fig.~\ref{3} we present time dependence of the electric charge in the S-layer for the stack with 10 JJ
combined with the outermost branch of CVC (the corresponding axis are shown by arrows). We found that the bias
current interval where the growing region (the beginning of that region is demonstrated in the inset (b)) of the
electric charge in S-layers is observed, is shorter now in compare with N=9 case, where the LPW with $k=
8\pi/9d$ is created at the same values of $\alpha$ and $\beta$. Compare this figure with Fig.~\ref{2}, we can
see that the amplitude of the charge is larger for the stack with even number of JJ. The inset (a) illustrates
the charge distribution among the layers and confirms the $\pi$-mode of LPW.
\begin{figure}[ht]
 \centering
\includegraphics[height=75mm]{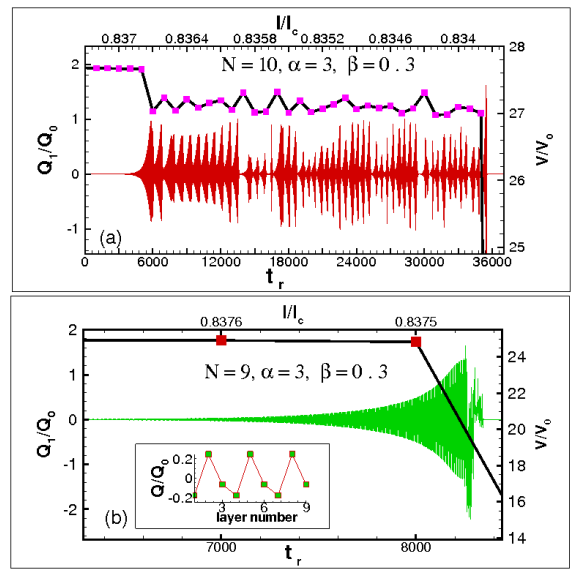}
\caption{Charge-time dependence and CVC at $\alpha= 3$, $\beta= 0.3$ and periodic BC: (a) for a stack with 10
JJ; (b) for a stack with 9 JJ. The inset shows the charge distribution among the layers corresponding to the
$2\pi/3d$-mode in the stack with 9 JJ. }
 \label{4}
\end{figure}

As we mentioned above, the wavelength of the LPW depends on the values of dissipation and coupling
parameters.\cite{sm-prl07} So, we can compare the stacks with 10 and 9 JJ  when another LPW are created and test
the idea concerning  the maximal amplitude of charge oscillations. In Fig.~\ref{4} the time dependence of the
charge in the first S-layer at $\alpha= 3$, $\beta= 0.3$ and periodic BC combined with the CVC of the outermost
branch is presented.

Fig.~\ref{4}(a) shows this dependence for the stack with 10 junctions. At chosen parameters the LPW with $k=
3\pi/5d$ is created. In case $N=9$ (Fig.~\ref{4}(b)) the inset illustrates the charge distribution among the
layers corresponding to the ($ 2\pi/3d$)-mode ($\lambda=3d$). Also we can see that the charge value on the
S-layers  in $k=2\pi/3d$ case is larger than in case of $k= 3\pi/5d$, which is the same result as we got before.
In addition to that, we tested the cases for $\lambda=4d$ and $\lambda=5d$ ( not presented here) and they
supported our idea. So, we may conclude that at fixed $\alpha$ and $\beta$ \emph{the charge value in S-layers is
larger for the stacks with "pure" parametric resonance} where the LPW with $\lambda= nd$ is created.
\begin{figure}[ht]
 \centering
 \includegraphics[height=55mm]{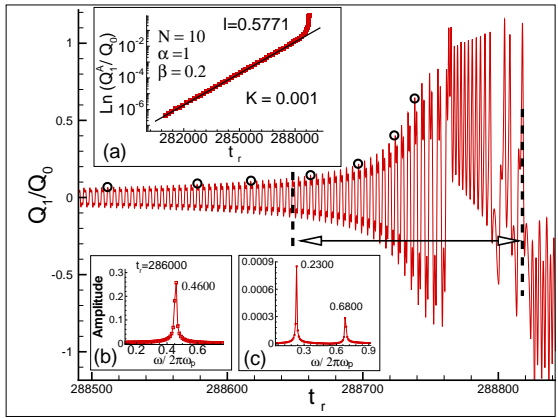}
\caption{Demonstration of the transition part (shown by double arrow) in charge-time dependence for the stack
with 10 JJ. Inset (a) shows the amplitude of charge oscillations in the logarithmic scale. Insets (b) and (c)
demonstrates the results of FFT analysis of voltage $V(t)$ and charge $Q_1(t)/Q_0$ time dependence in the
exponential growth part. }
 \label{5}
\end{figure}
\begin{figure}[ht]
 \centering
 \includegraphics[height=50mm]{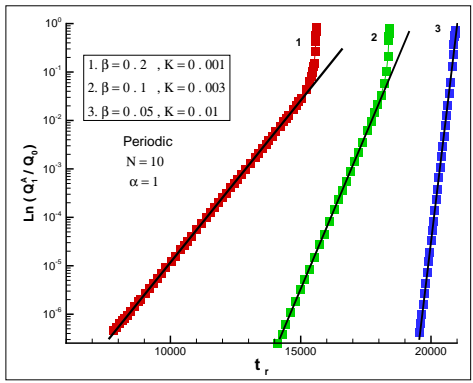}
\caption{Influence of dissipation magnitude on time dependence of the charge oscillation amplitude in the
logarithmic scale for the stack with 10 junctions at $\alpha= 1$.}
 \label{6}
\end{figure}

To demonstrate the character of the charge amplitude increasing in the growing region, we enlarged in
Fig.~\ref{5} the charge-time dependence for the stack with 10 JJ at $\alpha= 1$ and $\beta$ = 0.2. In inset (a)
we present the time dependence of the amplitude of $Q^A/Q_0$ in the logarithmic scale. The values of amplitude
were taken arbitrary at some time moments in total growing region (examples are shown by circles). We found two
parts in this dependence: exponential part and transition part (marked by double arrow) before jump to another
branch. In transition part the amplitude demonstrates a sharp increase  in short time interval in compare with
exponential part.  Insets (b) and (c) show results of FFT analysis of voltage $V(t)$ and charge $Q_1(t)/Q_0$
time dependence in the exponential grows part. They prove the parametric resonance condition $\omega_J=2
\omega_{LPW}$. In transition part this condition is broken. Writing the expression for electric charge  by
$Q_l/Q_0=\exp(Kt)$, we find $K=0.001$.

The Fig.~\ref{6} illustrates the influence of the dissipation magnitude ($\beta$ = 0.2, 0.1, and 0.05) on time
dependence of the charge oscillation amplitude in the logarithmic scale for the stack with 10 junctions at
$\alpha= 1$. From this figure we note the following features which are observed with decrease in $\beta$
(increase the McCumber parameter): i) the growing region is getting shorter; ii) the width of transition part is
decreasing; iii) the coefficient of the exponential growth $K$ is increased. We come to the important conclusion
that \emph{the parametric resonance features depend strongly on dissipation in the system}. The width of the
growing region is inversely proportional to the coefficient $K$. The value of $K$ is determined by the wave
number of LPW created at resonance. For all investigated  stacks with even number of JJ (in our simulations we
checked the stacks with $N$ in the interval (6,14)) at $\alpha= 1$, $\beta= 0.2$ and periodic BC we obtained the
same value of $K=0.001$.

In the supplement to this letter we include the avi files which present the animation of the charge dynamics in
the different time moments in the growing region. These animation files demonstrate the standing $\pi-mode$ of
created LPW (the charge on the nearest neighbor layers has the same value and opposite sign) in the exponential
part and its modification in the transition region.

As a summary, a manifestation of a novel type of hysteresis related to the parametric resonance in the system of
coupled Josephson junctions is demonstrated. The width of this hysteresis is inversely proportional to the
McCumber parameter and depends on coupling between junctions and the boundary conditions. The origin of this
hysteresis is related to the different charge dynamics for increasing and decreasing bias current processes. We
consider that these features are common for the systems demonstrating the parametric resonance. These features
can be used to develop new methods for determination of coupling and dissipation parameters of the system. We
show that the maximal value of electric charge amplitude realized in superconducting layers at the resonance
depends on the wavelength of the created LPW. A strong effect of the dissipation in the system on the width of
the parametric resonance is demonstrated.

We thank I. Rahmonov, M. Hamdipour, H. El Samman, S. Maize, M. Elhofy and Kh.Hegab for helpful discussions and
support this work.


\begin{thebibliography}{10}
\bibitem{mccumber68} D. E. McCumber, J.Appl.Phys. 39, 3113 (1968).
\bibitem{steward68} W. C. Steward, Appl.Phys.Lett. {\bf12}, 277 (1968).
\bibitem{sm-prl07} Yu. M. Shukrinov and F. Mahfouzi, Phys. Rev. Lett. 98, 157001 (2007).
\bibitem{Ozyuzer} L. Ozyuzer et al., Science 318, 1291 (2007).
\bibitem{koyama09} T. Koyama, H. Matsumoto, M. Machida, and K. Kadowaki,
     Phys. Rev. B 79, 104522 (2009).
\bibitem{ryndyk98} D. A. Ryndyk, Phys. Rev. Lett. {\bf80}, 3376 (1998).
\bibitem{koyama96} T. Koyama and M. Tachiki, Phys. Rev. B {\bf 54},  16183  (1996).
\bibitem{sms-physC06} Yu. M. Shukrinov, F. Mahfouzi, and P. Seidel, Physica C 449, 62 (2006).
\bibitem{matsumoto99} H. Matsumoto, S. Sakamoto, F. Wajima, T. Koyama, M. Machida, Phys. Rev. B {\bf 60}, 3666 (1999).
\bibitem{sm-physC06} Yu. M. Shukrinov and F. Mahfouzi, Physica C 434 (2006) 6-12.
\bibitem{sm-sust07} Yu. M. Shukrinov and F. Mahfouzi, Supercond. Sci. Technol. 20, S38 (2007).
\bibitem{sms-prb08}Yu. M. Shukrinov, F. Mahfouzi, M. Suzuki, Phys. Rev. B {\bf 78}, 134521 (2008).


\end{thebibliography}
\end{document}